# Numerical and experimental study of tonal noise sources at the outlet of an isolated centrifugal fan

**Martin Ottersten[1]**

Swegon Operation, Box 336, SE-401 25 Gothenburg, Sweden

Chalmers University of Technology, Department of Mechanics and Maritime Science,

Division of Fluid Dynamics, SE-412 96 Gothenburg, Sweden

martin.ottersten@chalmers.se

**Hua-Dong Yao**

Chalmers University of Technology, Department of Mechanics and Maritime Science,

Division of Fluid Dynamics, SE-412 96 Gothenburg, Sweden

huadong@chalmers.se

**Lars Davidson**

Chalmers University of Technology, Department of Mechanics and Maritime Science,

Division of Fluid Dynamics, SE-412 96 Gothenburg, Sweden

lars.davidson@chalmers.se

---

[1] Corresponding author.

**ABSTRACT**

*In this study, tonal noise produced by an isolated centrifugal fan is investigated using unsteady Reynolds-averaged Navier-Stokes (URANS) equations. This type of fans is used in ventilation systems. As the fan propagates tonal noise in the system, it can severely affect the life quality of people that reside in the buildings. Our simulation shows that turbulence kinetic energy (TKE) is unevenly distributed around the rotation axis. Large TKE exists near the shroud at the pressure sides of the blades. It is caused by the recirculating flow. Moreover, the position of the largest TKE periodically varies among the blades. The period corresponds to approximately 4 times the fan rotation period, it was also found in acoustic measurements. The magnitude of the tonal noise at the blade passing frequencies agrees well with experimental data. By analyzing the wall-pressure fluctuations, it is found that the recirculating flow regions with large TKE are dominant sources of the tonal noise.*

**INTRODUCTION**

In residential buildings, the systems of heating, ventilating and air conditioning (HVAC) are the sources of background noise. The level of noise annoyance in buildings can be evaluated in terms of loudness and spectral characteristics [1]. Exposure to tonal noise for a long-term can affect the autonomous and hormonal systems in the human body, leading to diseases such as high blood pressure, hearing loss, cardiac arrest, and mental disorders (aggressiveness and mood swings) [2, 3]. Therefore, it is of interest for HVAC system manufacturers to decrease the tonal noise.

HVAC systems are today driven by low speed isolated centrifugal fans, which means a fan with no diffusers, guide vanes and volutes. The flow upstream the fan is uniform and undisturbed. The blades are not interacting with any obstacle during the rotation. There are several studies on tonal noise produced by centrifugal fans, most of them are done on fans surrounded by a volute. Among others, Ref. [4] reports that the tonal noise at the blade passing frequencies (BPF) is caused by the interaction between the blades and the volute cut-off and that the amplitude can be affected by increasing the cutoff clearance. Therefore, isolated centrifugal fans should not exhibit tonal noise at BPF. In one of the first studies on isolated centrifugal fans [6], a tonal noise at BPF was presented. It was found that for this type of fans tonal noise at BPF is generated as a helical unsteady inlet vortex that interacts with the rotating blades near the fan backplate. In a study on nonuniform flow at the fan inlet [7], it was reported that the dominant BPF tonal noise is caused by strong flow distortion at the fan inlet and flow separation at the blade foot. To reduce the BPF tonal noise, a filter was assembled in front of the inlet [8]. This filter was found to be effective for straightening the inflow. The noise was consequently reduced due to the straightened inflow. Another way of decreasing the BPF tonal noise is to set flow obstructions upstream of centrifugal fans [9]. It was found that the BPF tonal noise is reduced by the obstructions with specifically designed shapes and locations.

In addition to fan inlets, flow structures at fan outlets were investigated in experimental measurements [10]. Vortices with high turbulence kinetic energy (TKE) are found at the pressure side of the blades between the trailing edge and the shroud. The same phenomenon was also observed and reported in another study on wake formation in a centrifugal fan [11]. In this study, they found a secondary flow on the pressure side of the blade, that interacts with the passage flow in the wake region near the shroud. However, the effect of these high TKE vortices on the tonal noise generation has not been addressed in the literature.

A convenient way of studying the tonal noise is to use numerical simulations. The Unsteady Reynolds averaged Navier-Stokes (URANS) method can simulate characteristic unsteady structures, which are responsible for tonal noise generation, with low computational costs. Its drawback is that it cannot provide the fluctuations that are important for the broadband noise generation. In a study on the isolated centrifugal fan noise [12], URANS equations were evaluated. The work shows good prediction of tonal noise magnitudes. The tonal noise in a cooling fan was effectively predicted in [13] with the URANS method coupled with the Ffowcs-Williams and Hawkings analogy. In [14] an acoustic prediction of a low-speed radial fan was performed. It was based on the URANS methodology and the FW-H acoustic analogy. The sound pressure level at the BPF was clearly captured. Also, URANS was used in [15]. The numerical results showed good agreement with experimental results.

This study aims to investigate the BPF tonal noise sources using URANS. The connection between TKE and the flow passing through the gap will be clarified. The influence of the recirculating regions on the surface pressure magnitudes on the blades will be reported. The relationship between the TKE and the wall-pressure fluctuations will be addressed. The calculated magnitudes and frequencies for the BPF tonal noise will be compared with experimental data. Finally, tonal noise sources at the fan outlet will be addressed.

**NOMENCLATURE**

| | |
|---|---|
| $d_1$ | Intake diameter |
| $d_2$ | Fan diameter |
| $d_3$ | Inlet diameter |
| $d_4$ | Outlet diameter |
| $b_2$ | Fan width at d2 |
| $h_1$ | The distance between inlet and fan |
| $h_2$ | The distance between outlet and fan |
| $w$ | Gap width |
| $z$ | Number of blades |
| $n$ | Rotation speed |
| $n_f$ | Rotation speed frequency |
| $BPF_0$ | Blade passing frequency |
| $BPF_1$ | First harmonic blade passing frequency |
| $p'$ | Sound pressure at the far-field |
| $c_0$ | Far-field sound speed |
| $T_{ij}$ | Lighthill stress tensor |
| $\rho$ | air density |
| $H(f)$ | Heaviside function |
| $P_{ij}$ | Compressive stress tensor |
| $n_i$ | Unit normal vector pointing |
| $\delta(f)$ | Dirac delta function |
| PSD | Power spectral density |

**CONFIGURATION**

The isolated centrifugal fan used in this study is illustrated in Figure 1. The parameters of the fan are specified in Table 1. The numerical setup is in line with the experimental setup. The microphones, M1 and M2, are placed both upstream and downstream of the fan, respectively. There is a gap between the stationary inlet duct and the rotating fan shroud. Air can flow into the fan through the gap. The flow at the inlet is undisturbed and uniform.

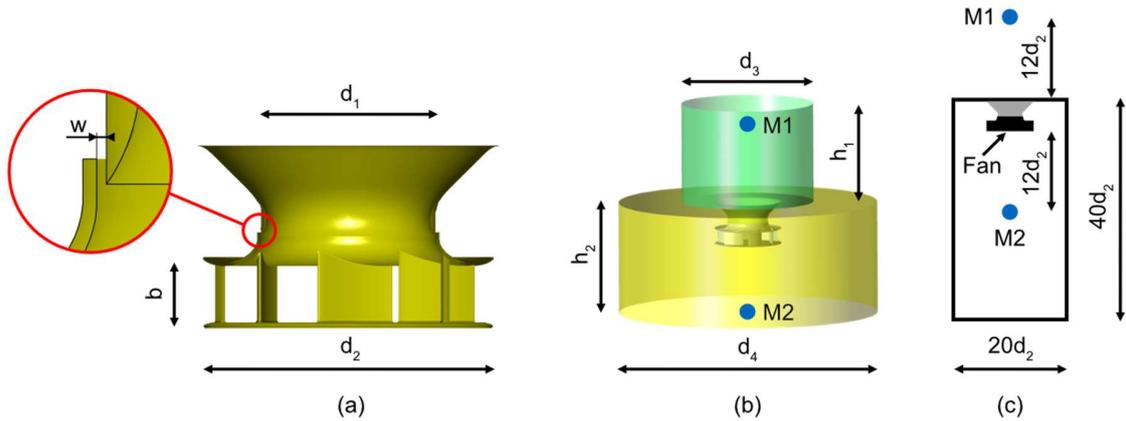

*Figure 1. a) The fan geometry b) the computational domain c) the experimental setup, where the microphones are indicated*

*Table 1. Parameters of the fan*

| Intake diameter | $d_1$ | 0.165 m |
|---|---|---|
| Fan diameter | $d_2$ | 0.268 m |
| Inlet diameter | $d_3$ | 0.6 m |
| Outlet diameter | $d_4$ | 1.1 m |
| Fan width | b | 0.053 m |
| The distance between inlet and fan | $h_1$ | 0.4 m |
| The distance between outlet and fan | $h_2$ | 0.5 m |
| Gap width | w | 0.0015 m |
| Number of blades | z | 7 |
| Rotation speed | n | 2800 rpm |
| Rotation speed frequency | $n_f$ | 46.7 Hz |
| BPF | $BPF_0$ | 326.7 Hz |
| First harmonic frequency | $BPF_1$ | 653.4 Hz |

**Experimental setup**

The sound pressure is measured in the test rig that is a reverberation room. The geometry of the experimental setup is larger than that for the HVAC, in order to minimize the effect of the boundary conditions at the walls in the surrounding chamber, Figure 1. The microphones are PCB Piezotronics 130A20 40791. National Instruments program Sound and vibration was used to sample the data. The power spectrum averaging mode was selected to RMS averaging with the average number set to 3.

**NUMERICAL METHODOLOGY**

**The unsteady RANS**

The flow is incompressible. This assumption is made since the Mach number is less than 0.3 [16, 17]. The Mach number based on the blade tip velocity is 0.1. An incompressible URANS solver is applied to the unsteady simulation. The simulation is performed using the software Fluent 19.0 [18].

The turbulence model is the k-ω shear-stress transport (SST), which was used in [13, 19, 20]. The segregated flow solver is used to solve the discretized equations. The pressure-velocity coupling is calculated through the algorithm SIMPLEC (Semi-Implicit Method for Pressure-Linked Equations-Consistent). The SIMPLEC procedure is similar to the SIMPLE procedure. The only difference lies in the expression used for the face flux correction. The modified correction equation has been shown to accelerate the convergence for the problems where the pressure-velocity coupling is the main deterrent for obtaining a solution [21]. The discretization scheme for the pressure is second-order. The momentum and turbulence kinetic energy (TKE) are discretized with a second-order upwind scheme. The Kato-Launder formulation and production limiter are applied to avoid the excessive generation of the turbulence energy near stagnation points.

A bounded second-order implicit method [21] is used to discretize the time derivative. The time step is set to $\Delta t = 5.95 \cdot 10^{-5}$ s, corresponding to 1° rotation of the fan region. Local Courant numbers in most of the regions in the computational domain are smaller than 1, while the maximum Courant number is 66 in small regions near the tip. The maximum number of inner iterations at every time step is 20.

The coefficients for the SST model are the same as those for the standard k-ω model. The default values of the coefficients in the software are adopted. For the SST k-ω model, σ$_{k,1}$ = 1.176, σ$_{k,1}$ = 1.0, σ$_{ω,1}$ = 2.0, σ$_{ω,2}$ = 1.168, a$_1$ = 0.31, β$_{i,1}$ = 0.075 and β$_{i,1}$ = 0.0828.

**The FW-H acoustic analogy**

The incompressible URANS solver gives only the hydrodynamic pressure. The noise, which accounts for the acoustic pressure, is excluded from the incompressible simulation. To predict the noise, the Ffowcs Williams and Hawkings (FW-H) acoustic analogy is coupled with the URANS. The integral surfaces for the FW-H method are set on the blades, the backplate, and the shroud (see Figure 2).

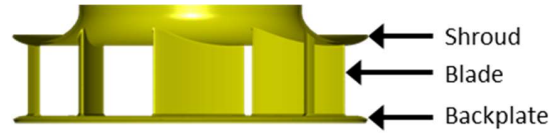

Figure 2. Illustration of the integral surfaces for the FW-H method

The FW-H equation is written as the following [21]:

$$\frac{1}{c_0^2}\frac{\partial^2 p'}{\partial t^2} - \frac{\partial^2 p'}{\partial x_i^2} = \frac{\partial[\rho_0 v_n \delta(f)]}{\partial t} - \frac{\partial[P_{ij}n_j \delta(f)]}{\partial x_i} + \frac{\partial^2[T_{ij}H(f)]}{\partial x_i \partial x_j} \quad (1)$$

where T$_{ij}$ and P$_{ij}$ are

$$T_{ij} = \rho v_i v_j + P_{ij} - c_0^2(\rho - \rho_0)\delta_{ij} \quad (2)$$

$$P_{ij} = p\delta_{ij} - \mu\left[\frac{\partial v_i}{\partial x_j} + \frac{\partial v_j}{\partial x_i}\right] \quad (3)$$

The right-hand side of Equation (1) represents three types of sources. The first, second and third terms represent monopole, dipole and quadrupole, respectively. It was shown in [22] that the dipole is the dominant source of fan noise at low Mach number. Hence, the dipole term is the only one used in this study.

A body can be regarded to be acoustically compact when its characteristic dimension is small compared to the wavelengths of the sound waves produced by it. The fan radius is, $a = 0.134\ m$. The characteristic wavelength of the tonal noise is, $\lambda = 0.52\ m$, which is calculated by dividing the speed of sound with the $BPF_1$. Due to $a \ll \lambda$ the fan can be treated as an acoustically compact body [23] with respect to the $BPF_1$. The ambient fluid for the FW-H method is set with the density of 1.225 kg/m³ and, the speed of sound of 340 m/s. The reference acoustic pressure is 2·10⁻⁵ Pa.

**NUMERICAL SETTINGS**

The computational domain is illustrated in Figure 3. The entire computational domain is divided into stationary and rotating domains. The domains are connected through an nonconformable interface [21]. The fan-region is contained in the rotating domain, in which mesh cells move with the rotating fan.

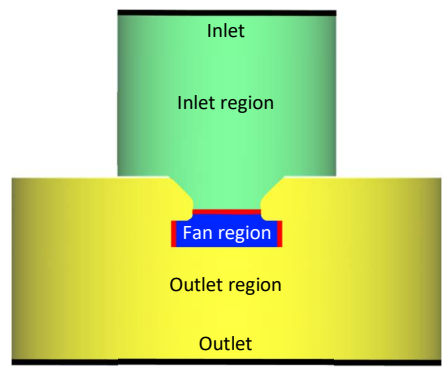

*Figure 3. The fan region (blue) is a rotating domain. The inlet region (green) and the outlet region (yellow) are non-rotating domains. The interfaces between the rotating and non-rotating domains are marked with red.*

The tone frequencies are $BPF_0$=326.7 Hz and $BPF_1$=653.4 Hz. The sampling frequency is equal to $f_s = \Delta t \cdot k$. The data are recorded every k time step, where k = 10. The selected sampling frequency determines the maximum resolved frequency as $f_s/2$ according to the Nyquist law. The lowest resolved frequency is $\Delta f = 200\ Hz$. This condition is satisfied by setting the recording time of signals to $T = 20/\Delta f = 0.2\ s$, which is approximately 10 rotation periods.

The mass-flow boundary condition is set at the inlet. The modeled turbulence intensity is set to I=4 %, according to $I = 0.16(R_e)^{-1/8}$ [21]. Here Re is computed based on the inlet diameter and the streamwise velocity at the inlet. The modeled turbulence length scale is set to ℓ=0.05 m based on $\ell = 0.7 d_{inlet}$, where $d_{inlet}$ is the diameter of the inlet duct. The pressure-outlet boundary condition with the Gauge pressure of 0 Pa is set at the outlet. The no-slip boundary condition is used on the walls.

The under-relaxation factors in the segregated flow solver are set to 0.7 for the momentum, 0.3 for the pressure, and 0.8 for the TKE. The current numerical method was validated in [12]. In that study, the results for the fan aerodynamic properties are consistent with the experimental data.

**COMPUTATIONAL MESH**

The mesh contains prism layers near the walls and polyhedral elements in the rest of the computation domain. The polyhedral mesh generation method can control the element growth ratio to give smooth changes of element size. The growth rate is set to 1.05, as suggested in previous studies [24, 25].

The mesh density study is performed with three different mesh sizes, named Mesh A, Mesh B, and Mesh C. The same strategy was used on all three. The mesh density is scaled with 0.85 for Mesh A and with 1.2 for Mesh C, compared with Mesh B. This is done by changing the element size on the walls and in the volume in the fan region. The properties of the meshes are listed in Table 3 and are illustrated in Figure 4.
For Mesh B the $y^+$ near the fan surfaces are less than 1, except for a small region near the leading edge of the blades. Mesh B is illustrated in Figure 5. The surface element sizes of Mesh B in the fan are presented in Table 3.

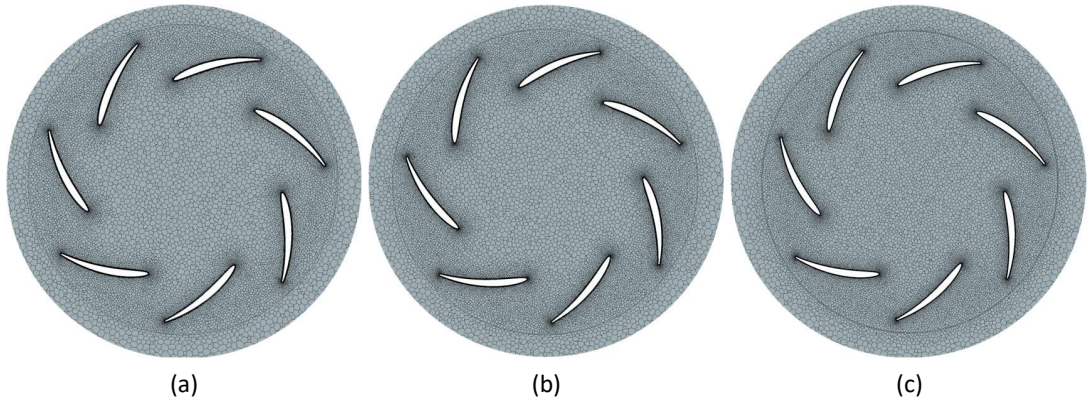

*Figure 4. Volume mesh in a plane normal to the rotation axis in a) Mesh A, b) Mesh B, and c) Mesh C*

*Table 2. Mesh summary*

|  | Nodes | Cells |
|---|---|---|
| **Mesh A** | 12 041 059 | 7 112 081 |
| **Mesh B** | 16 132 851 | 8 366 763 |
| **Mesh C** | 18 108 853 | 9 994 218 |

*Table 3. Surface element sizes for Mesh B*

|  | Maximum cell size [$m^3$] | Minimum cell size [$m^3$] |
|---|---|---|
| **Blade** | $8.2 \cdot 10^{-11}$ | $7.6 \cdot 10^{-13}$ |
| **Shroud** | $1.4 \cdot 10^{-10}$ | $1.1 \cdot 10^{-12}$ |
| **Backplate** | $4.0 \cdot 10^{-9}$ | $2.3 \cdot 10^{-12}$ |
| **Fan (rotation region)** | $1.2 \cdot 10^{-7}$ | $5.1 \cdot 10^{-14}$ |

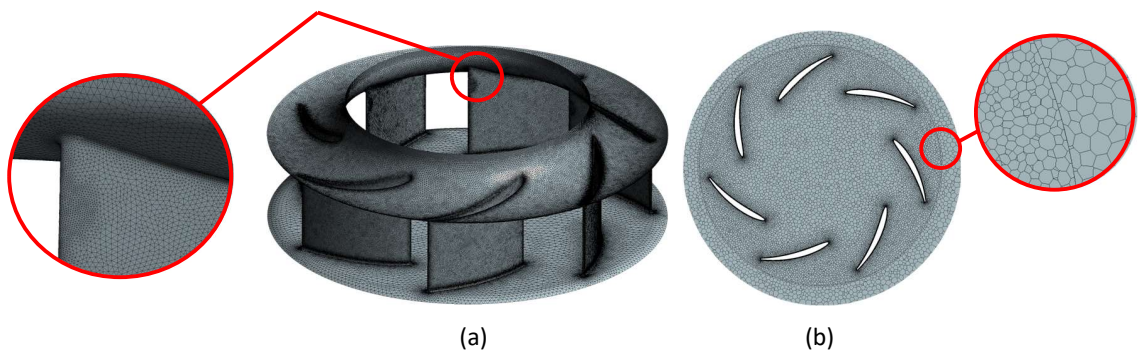

*Figure 5. a) Surface mesh on the fan for Mesh B. b) volume mesh in a plane normal to the rotation axis*

## RESULTS AND DISCUSSION

The cut planes used to visualize flow variables in the subsequent discussion are illustrated in Figure 6. They were chosen so that the flow behavior could be studied close to the shroud, in the middle between the shroud and the backplate, and close to the backplate. In addition, the blades are numbered.

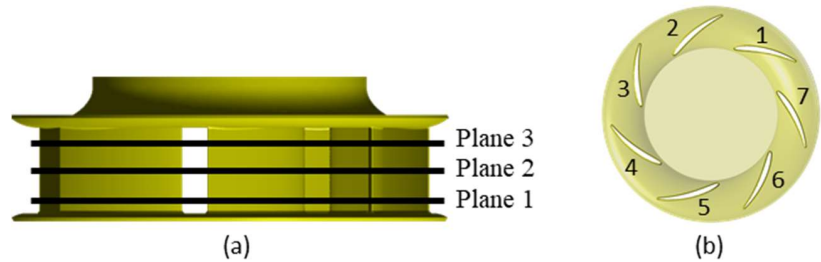

*Figure 6. a) The cut planes used to investigate the flow behavior. Plane 1 is located at $0.07 \cdot b_2$ from the backplate, Plane 2 at $0.47 \cdot b_2$, Plane 3 at $0.66 \cdot b_2$. $b_2$ is defined in Figure 1. b) Blade location at the arbitrary time $t=t_0$*

The important physical data for the fan was compared between the different mesh densities and the experiment in Table 4. The results from Mesh B and Mesh C agree well with experimental data. For Mesh A the agreement was poorer, because the mesh was too coarse. Mesh B was selected for the acoustic calculation, in order to save computational costs.

*Table 4. Fan performance data for the mesh density study*

|  | Static pressure [Pa] | Torque [Nm] | Volume flow [m³/s] |
|---|---|---|---|
| **URANS - Mesh A** | 311 | 1.200 | 0.395 |
| **URANS - Mesh B** | 306 | 1.215 | 0.395 |
| **URANS - Mesh C** | 305 | 1.224 | 0.395 |
| **Experimental** | 304 | 1.234 | 0.395 |

Fan performance is affected by flow separations on the blades [26]. The location and the size of the flow separations can be detected with the wall shear stress [27]. The wall shear stress and streamlines are illustrated in Figure 7a. Recirculating flows are observed between the shroud and the blade trailing edges. The recirculating flows enable the reduction of wall shear stress on the blades. In the recirculating regions, the velocity magnitudes are small. Meanwhile, a low-pressure region is observed upstream of the recirculating region.

A snapshot of the total pressure below -610 Pa and streamlines is illustrated in Figure 7. The same streamlines in subfigure a are shown in subfigure b. A low total pressure region is found at the blade leading edges. The size of the recirculating flow regions, which are indicated with streamlines, are unevenly distributed among the blades. Similar uneven distributions are observed for the velocity magnitudes and total pressure. The sizes of the intensive regions of these variables are associated with the recirculating region.

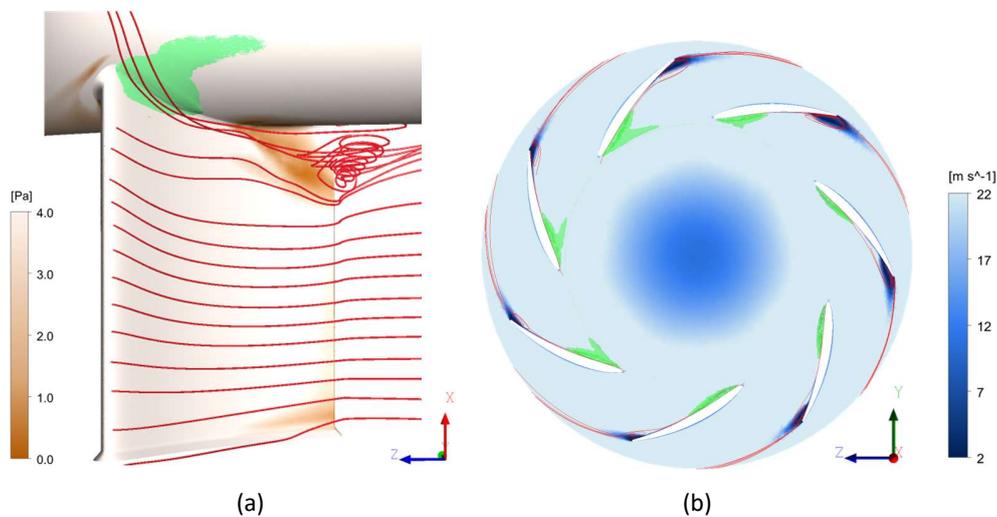

*Figure 7 Snapshots of a) contours of the wall shear stress on the walls and b) velocity magnitudes in Plane 3. Streamlines are visualized in red and contours of the total pressure at -610 Pa visualized in green. The location of Pane 3 is given in Figure 5a.*

Figure 8 shows streamlines near the shroud. They show that the recirculating region is formed by the flow passing the gap. This finding is consistent with the results reported by [26]. In their study, it was found that the gap produces trailing edge flow separation (as in Figure 8), which affects the fan performance.

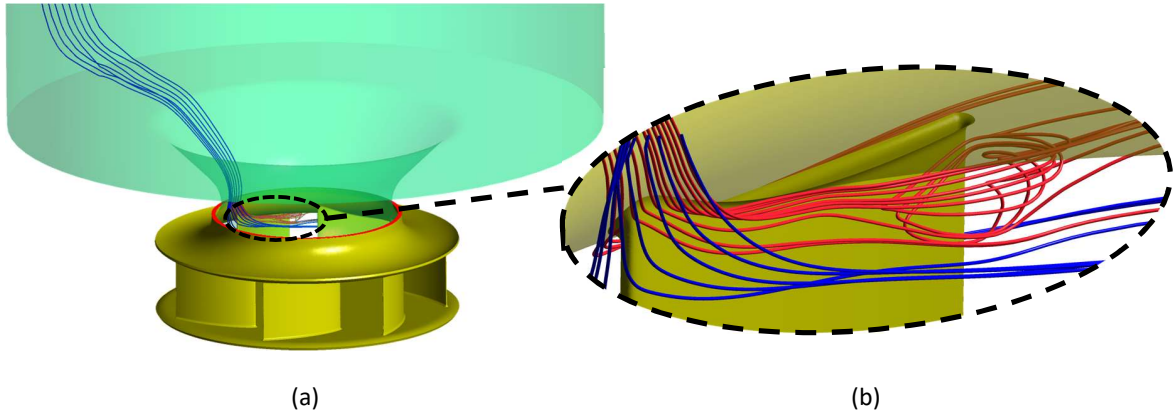

Figure 8. Streamlines streaming from the inlet (blue) and from the gap (red). The gap, w, is shown in Figure 1. a) Fan and inlet duct b) zoomed at blade and shroud intersection

Snapshots of the modeled TKE on the cut planes are illustrated in Figure 9. As shown by the TKE on Plane 3, the high TKE region is associated with the recirculating flow, see Figure 7b. Moreover, the regions with high TKE show uneven distribution among the blades as well. This is similar to the aforementioned variables such as the wall shear stress and total pressure, see Figure 7.

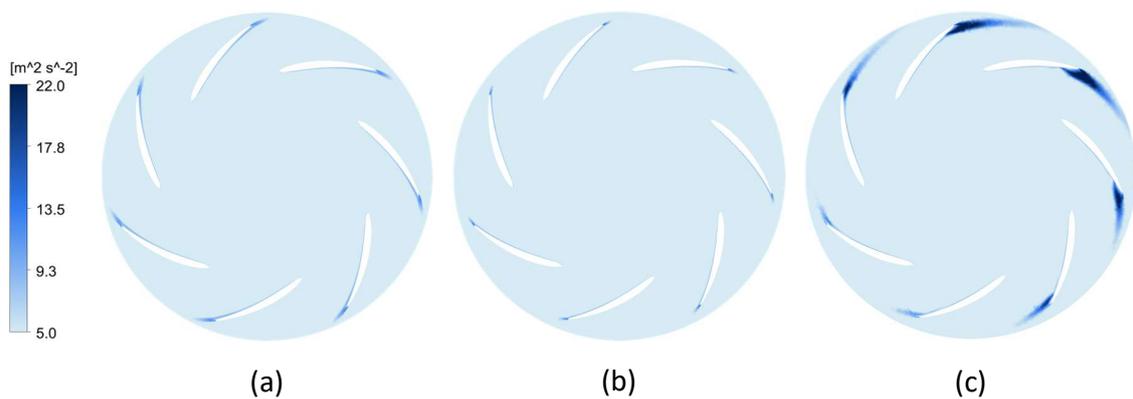

Figure 9. The modeled TKE in a) Plane 1, b) Plane 2, and c) Plane 3. Locations of Planes are given in Figure 5a.

Figure 10 shows the instantaneous surface pressure magnitudes along the intersecting lines between the blades and the cut planes shown in Figure 7. The pressure in Plane 1 is similar for all blades. In Plane 3, however, Blade 1 shows different pressure magnitudes as compared with the other blades. On the pressure-side, the largest pressure is observed near the leading edge of Blade 1. The pressure-side pressure on Blade 1 decays faster between 0.2-0.8 chord length than the other blades, whereas this decaying trend becomes slow above 0.8 chord length. The pressure-side pressure on Blade 1 is larger than on the other blades above 0.85 cord length. On the suction-side, the smallest pressure appears near the leading edge of Blade 1. The distinct behavior of Blade 1 is caused by the recirculating flow.

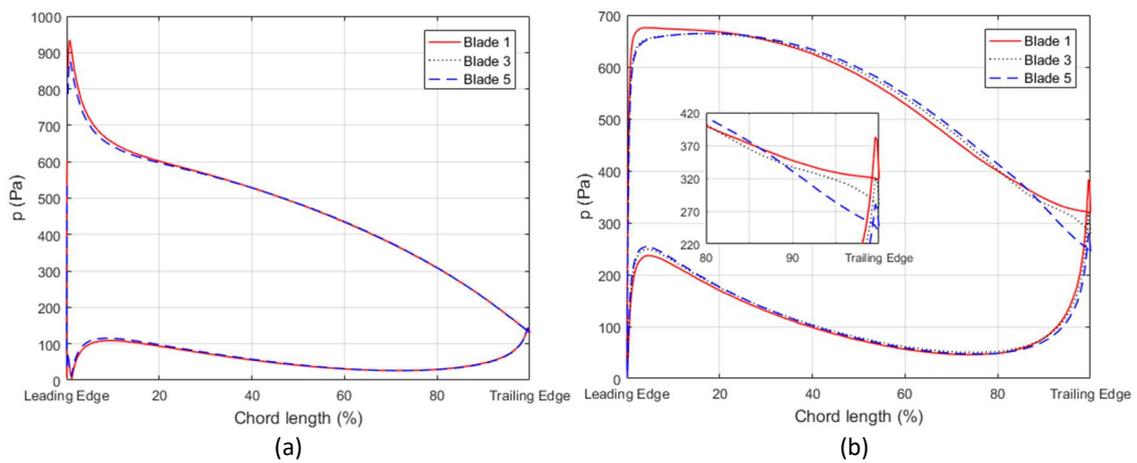

*Figure 10. The pressure distribution at $t=t_0$ at the intersection between the blade and a) Plane 1 b) Plane 3. The blade numbers are given in Figure 5b.*

Figure 11 shows snapshots of the unevenly-distributed TKE in Plane 3 and the velocity magnitudes in the gap, where T is the fan rotation period. It is observed that the largest TKE region moves among the blades in sequence. This region appears at the same blade, which is colored in green in the figure, at $t_0$ and $t_0+4T$. It is therefore concluded that the revolution period of the largest TKE region is 4T. The velocity magnitudes at the fan gap are also illustrated in Figure 11. The regions with high-velocity magnitudes always occur upstream of the blade with high TKE (the fan rotates in clockwise direction). This phenomenon can be explained based on streamlines illustrated in Figure 8. The streamlines passing through the fan gap become recirculating near the blade trailing edge. The same physical behavior was also observed in experimental measurements for centrifugal compressors [10]. The large TKE can be linked to a meridional curvature effect [11], which means that the flow direction is changed from axial to radial. The large TKE is also related to low total pressure between the shroud and the blade leading edge, as shown in Figure 7.

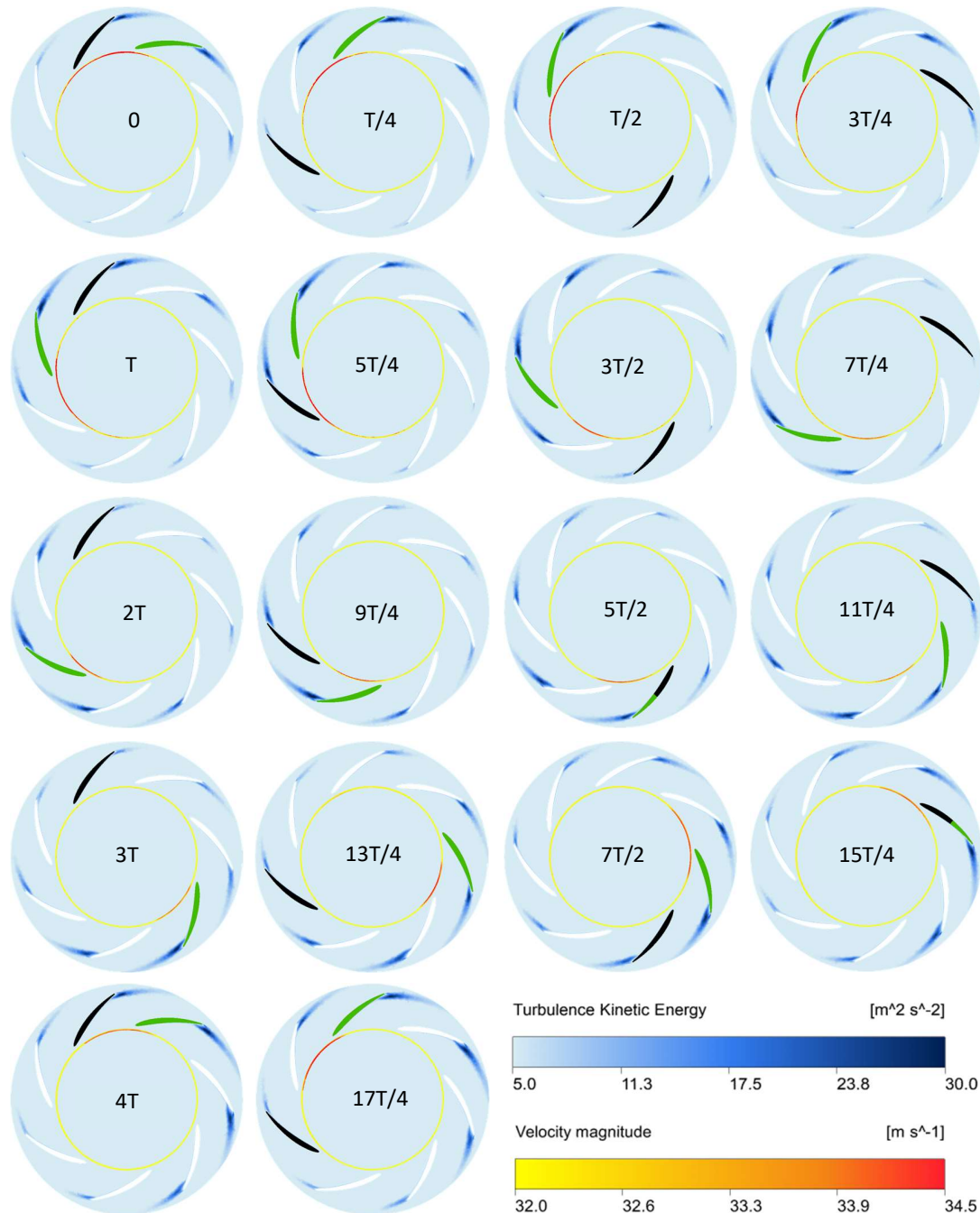

*Figure 11. Snapshots of velocity magnitudes through the gap and the turbulence kinetic energy in Plane 3. The blade colored in black displays the rotation of the fan. The blade colored in green indicates the region with the large TKE. The location of Pane 3 is given in Figure 5a.*

Figure 12 illustrates snapshots of the total pressure on the shroud. There are regions with low total pressure that move with the same angular speed as the high TKE and the high-velocity magnitude through the gap. It was reported in [27] that the pressure in the area of the gap fluctuates.

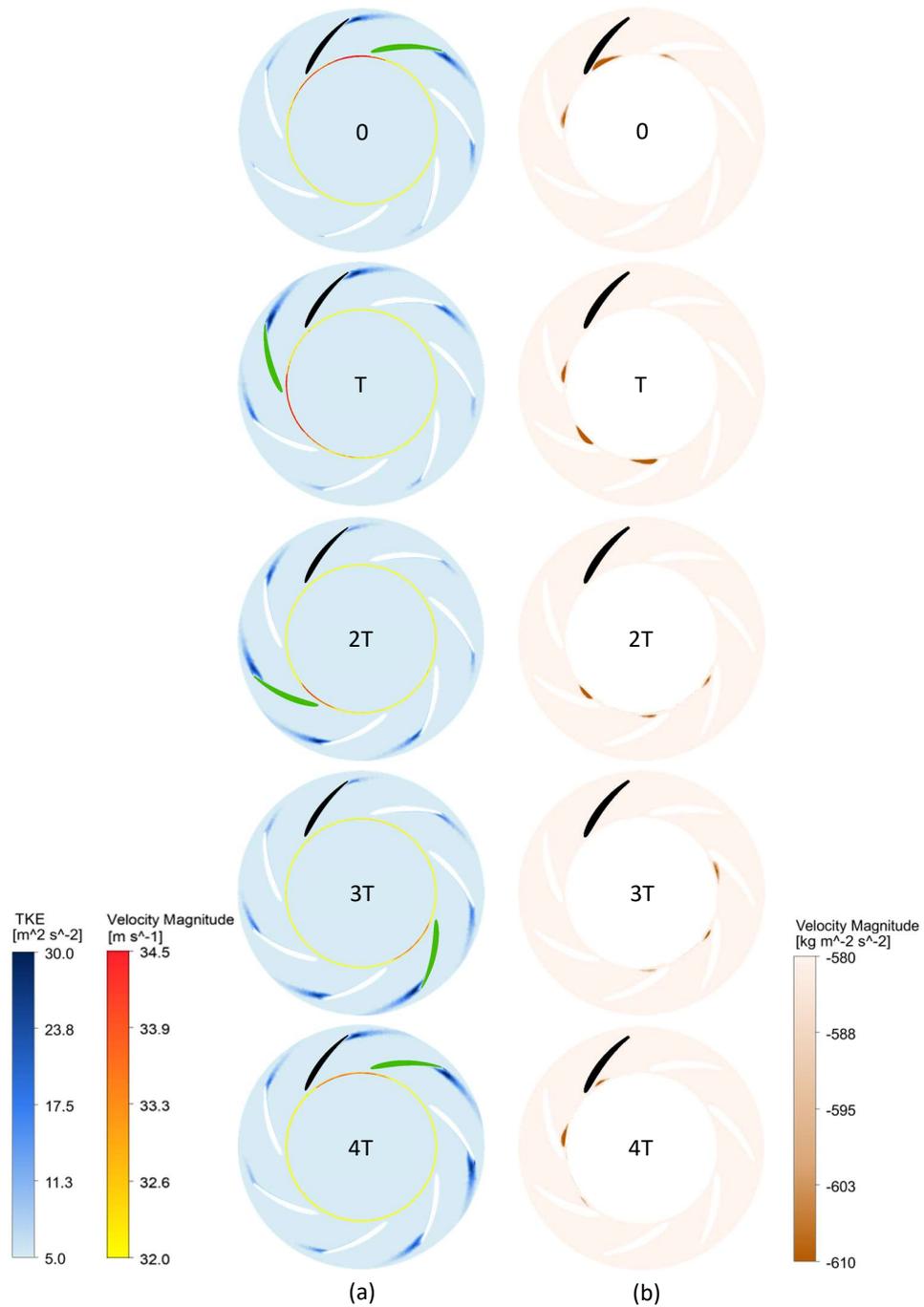

(a)    (b)

*Figure 12. Snapshots of a) the TKE in Plane 3 and the velocity magnitude through the gap, and b) the total pressure on the shroud. The blade colored in black displays the rotation of the fan. The blade colored in green indicates the region with large TKE. The location of Pane 3 is given in Figure 5a.*

The revolution period of high TKE (corresponding to the recirculation), velocity magnitudes, and the total pressure is 4T, as shown in Figures 11 and 12. This can also be shown with the pressure distribution on the blades with respect to time. This is illustrated at blade 7 in Figures 13a and 13b, at the location of the intersection of blade 7 and Plane 1 and Plane 3, respectively. The pressure distribution is constant during the fan revolutions at the location of Plane 1. But in Plane 3 it is not constant. At $t=t_0+4T$, the pressure distribution at the trailing edge is almost the same as that at $t_0$. The largest pressure difference is 70 Pa and occurs at 2T. Hence, the fan has rotated two revolutions and the high TKE has rotated half a revolution.

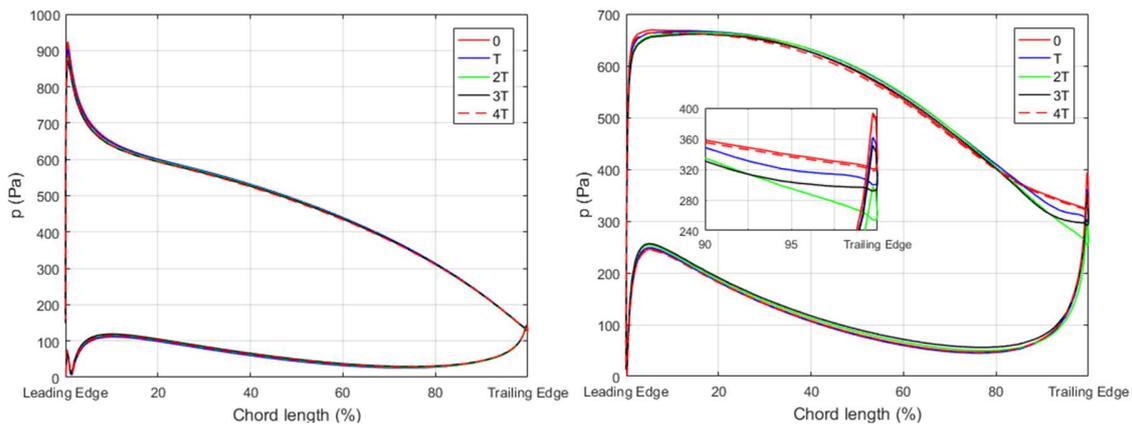

*Figure 13. The pressure distribution at different times at the intersection of blade 7 (see Figure 5b) and a) Plane 1 b) Plane 3. Location of Planes and blade numbers are given in Figure 5.*

It has been shown above that the regions with high TKE are caused by the separation on the blades. The separation is larger at low volume flow (with constant rpm) which was explained in [28]. The static pressure rise for the fan from the experiments with constant rotation speed (2800 rpm) is plotted in Figure 14. It can then be concluded that the separation is larger at operation point 1 than point 2. The operation point used in this study was point 2 because the amplitude for the $BPF_0$ was largest. The experimental measured power spectral density (PSD) upstream (M1) and downstream (M2) the fan for operation points 1 and 2, is presented in Figure 15. The revolution period of the region with high TKE is 4T, as concluded in the simulation. This revolution period can be seen as a tonal frequency at ¼ of the $BPF_0$ and ¼ of the rotation frequency $n_f$. This confirms the conclusion made from the simulations. The amplitude for this tonal frequencies is larger at point 1, which confirms that it increases with decreased volume flow (larger separation). These tonal frequencies cannot be explained by the rotating stall. Studies on rotating stall in centrifugal fans are numerous. Among others [29] showed that the typical tonal noise at the rotating stall is in the range of 60 to 80 % of BPF frequency.

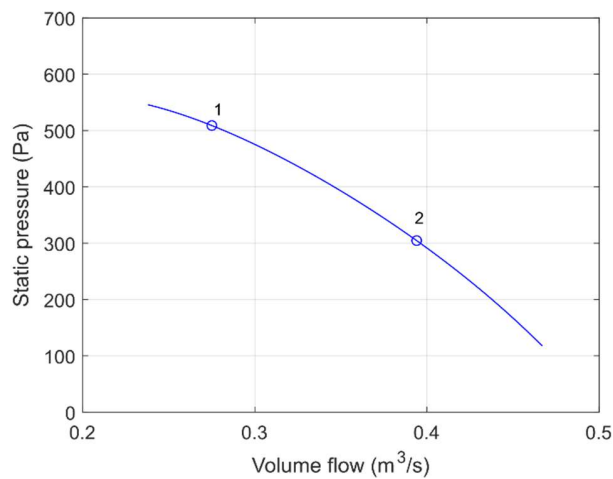

*Figure 14. The pressure fan curve measured in experiment, with constant rotation speed (2800 rpm)*

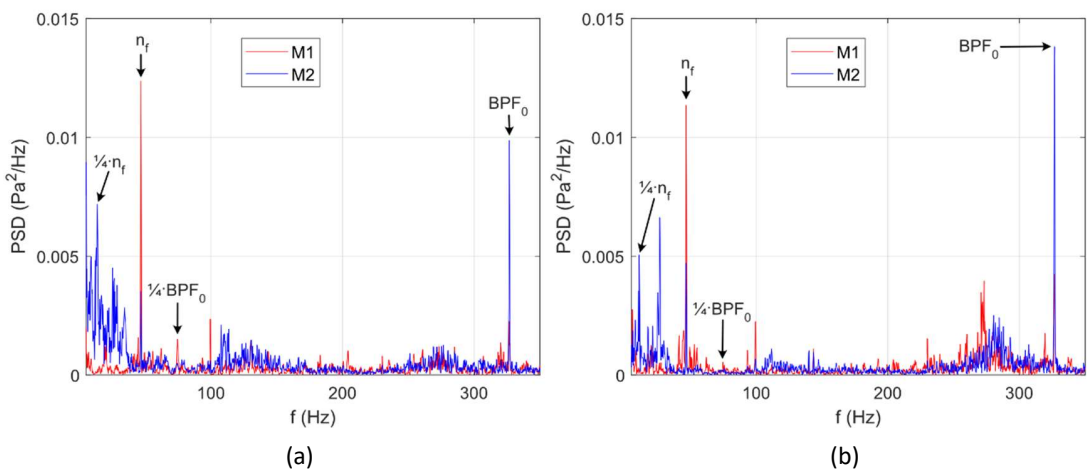

*Figure 15. PSD measured in experiment, upstream (M1) and downstream (M2) at operation point a) 1 and b) 2*

The PSD predicted at microphones, M1 near the inlet and M2 near the outlet are compared with experiments in Figure 16. The predicted tones at M1 ($BPF_0$=326.7 Hz and $BPF_1$=653.4 Hz) agree well with the experimental results in both magnitudes and frequencies. At M2, the tonal frequencies are captured, but the predicted magnitudes of the tones are different from the experimental results. A possible reason is that the interaction between the noise and the resonance modes of the fan affect the tonal noise magnitudes [7, 8]. This effect is more obvious near the outlet than near the inlet. The same phenomenon is found in the experimental results. The tonal frequencies ¼ $BPF_0$ and ¼ $n_f$ (showed in Figure 15) cannot be seen in the simulated result, because the lowest resolved frequency was 200 Hz (as described above).

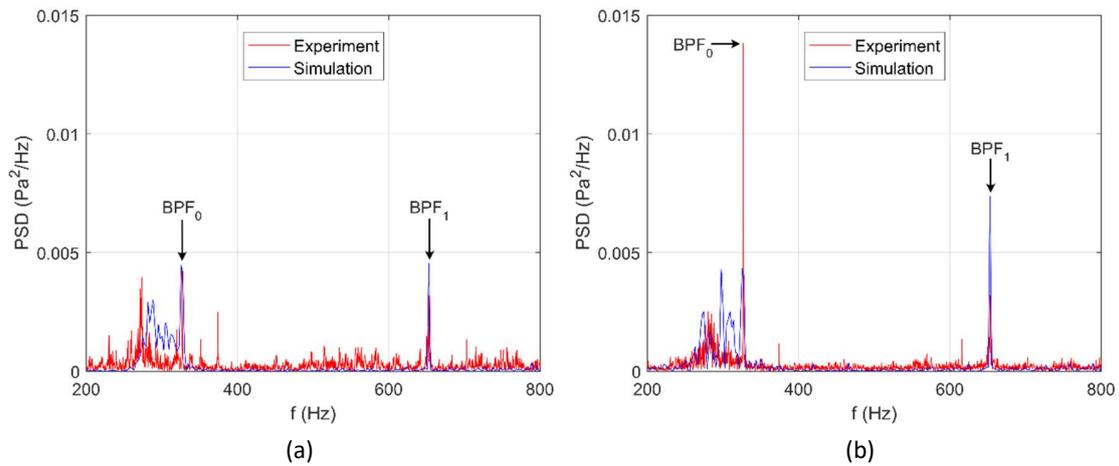

*Figure 16. Comparison of PSD at M1 and M2. ($BPF_0$=326.7 Hz and $BPF_1$=653.4 Hz)*

It is possible from the band filtered power spectral density of the wall-pressure fluctuations to provide an indication of the acoustic sources. The surface pressure levels at the $BPF_0$ (326.7 Hz) are illustrated in Figure 17. The highest-pressure fluctuation levels are located between the shroud and the trailing edge on the pressure side of the blade i.e. at the recirculating region (see Figure 8b). In this region, the pressure fluctuation levels on the shroud are high (see Figures 7b and 12b). The surface pressure levels at $BPF_1$ (653.4 Hz) are illustrated in Figure 18. The highest values are at the same location as for $BPF_0$ but with a lower magnitude. The pressure fluctuations are unevenly distributed among the blades. The uneven distribution agrees with the high TKE distribution (see Figures 10, 11 and 12).

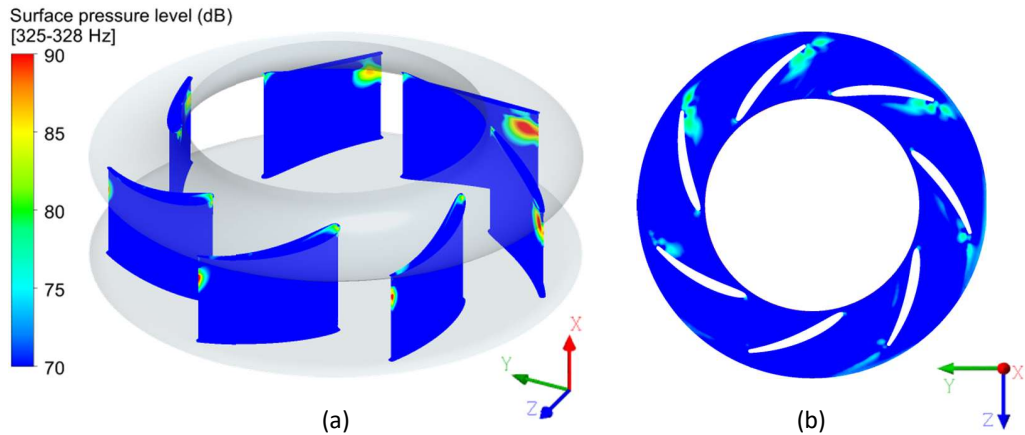

Figure 17. Surface pressure level at the $BPF_0$ (326,7 Hz) on a) the blades b) the shroud

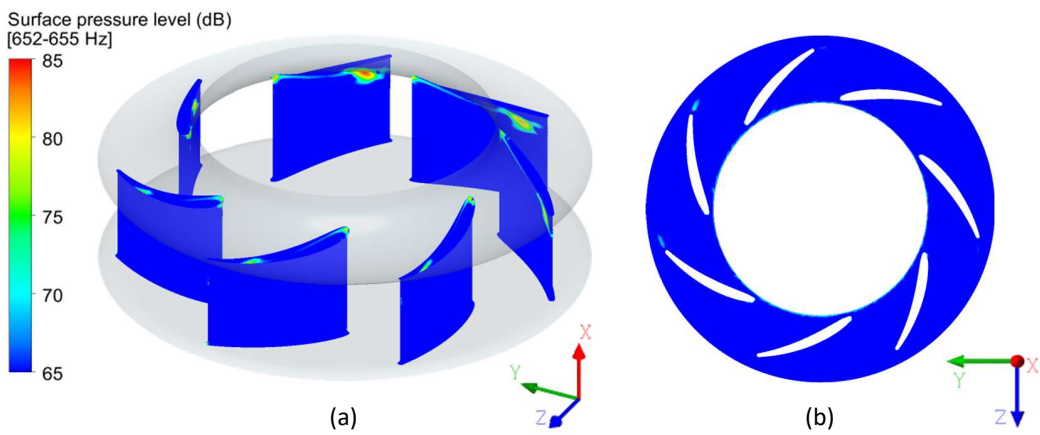

Figure 18. Surface pressure level at the $BPF_1$ (653.4 Hz) on a) the blades b) the shroud

**CONCLUSION**

Regions with high turbulence kinetic energy (TKE) are found between the shroud and the blade trailing edge, corresponding to the blade pressure side. These regions are connected to flow recirculations, which are caused by the flow passing the gap. The large TKE is unevenly distributed among the blades and periodically varies in time. The revolution period is approximately 4T, where T is the revolution period of the fan. The behavior of the velocity magnitudes in the gap and the total pressure on the shroud agree with that of the high TKE. Their revolution periods are also 4T. This revolution period is also found in acoustic measurements in the test rig, where tonal frequencies were found at ¼ $BPF_0$ and ¼ $n_f$.

The predicted tones at $BPF_0$ and $BPF_1$ near the inlet, are consistent with experimental data (frequencies and magnitudes). The predicted tonal magnitudes near the outlet, are slightly different from the experiments although the frequencies are captured. A possible reason is an interaction between the noise and the resonance modes of the fan in the experiments.

The wall-pressure fluctuations indicate that the locations of the acoustic sources at the $BPF_0$ and $BPF_1$ agree with the location of the recirculating region. The surface pressure levels at $BPF_0$ are larger than $BPF_1$. The regions with high surface pressure levels are unevenly distributed among the blades. The uneven distribution agrees with the regions with high TKE.

**ACKNOWLEDGMENTS**

The present work is financed by Swegon Operation.